\documentclass[english,10pt,pra,aps,showpacs,twocolumn,amsmath,amssymb,
longbibliography,
nofootinbib,
floats]{revtex4-2}
\usepackage[T1]{fontenc}
\usepackage[latin9]{inputenc}
\usepackage{verbatim}
\usepackage{refstyle}
\usepackage{amsmath}
\usepackage{amssymb}
\usepackage{graphicx}
\usepackage{color}
\usepackage{bm}
\usepackage{hyperref}
\usepackage{times}
\makeatletter

\usepackage{braket}

\usepackage{babel}
\begin{document}

\title{Adiabatic sweep theorem for three-dimensional dipolar Bose gases}
%from the variational theorem for the scattering length}

\author{Alexander Yu.~Cherny}
\email{cherny@theor.jinr.ru}
\affiliation{Bogoliubov Laboratory of Theoretical Physics, Joint Institute for Nuclear
Research, 141980, Dubna, Moscow region, Russia}

\date{\today}

\begin{abstract}
The variational theorem for the scattering length in the presence of the dipole-dipole interaction is developed. The theorem is applied to the spinless dipolar Bose gas in three dimensions. We calculated analytically the long-range tails of the single-particle momentum distribution and static structure factor, and the pair distribution function at short distances. The momentum distribution is inversely proportional to $q^4$ with the anisotropic prefactor. In the absence of the dipole-dipole interaction, Tan's adiabatic sweep theorem is reproduced as a particular case. For the homogeneous dilute Bose gas, all the relations are calculated analytically.
\end{abstract}
\maketitle

\section{Introduction}
\label{sec:intro}

Dipolar quantum gases are an important tool to study quantum many-body effects both theoretically and experimentally (see the reviews \cite{Baranov08,Lahaye09,Baranov12,Pitaevskii16:book}). The anisotropic long-range dipole-dipole interaction leads to a variety of their properties and provides a way of controlling them with external electromagnetic fields.

The physical properties of the dilute dipolar Bose gas with magnetic or electric moment are governed by two parameters:
%the low-energy scattering amplitude, which is proportional to
the scattering length $a$ and the absolute value of the dipole moment $d$. This property is called universality. The scattering length arises due to an additional short-range interaction (for details, see Sec.~\ref{sec:pseudo} below). It is convenient to introduce the effective dipole range $r_{\mathrm{dd}}=m d^2/(3\hbar^2)$, which has the dimension of length and determines the strength of the dipole-dipole interaction (see, e.g., Ch.~25 of Ref.~\cite{Pitaevskii16:book}).  The dimensionless parameter $\epsilon_{\mathrm{dd}}=r_{\mathrm{dd}}/a$ characterises the relative strength of the dipole-dipole interaction. Here $m$ is the mass of interacting particles.

The dipolar Bose gas in three dimensions can be thermodynamically unstable due to the attractive part of the dipole-dipole interactions. To stabilize it, the additional short-range interaction should be repulsive ($a>0$) and sufficiently large ($a>r_{\mathrm{dd}}$).  Otherwise, the system breaks down into tiny clusters called droplets (see the reviews \cite{Ferrier-Barbut19,Luo20,Bottcher20}) or even collapses completely. Recently, the ground state energy and correlation functions were obtained for the homogeneous dilute dipolar Bose gas \cite{cherny19b} when the stability conditions are satisfied.

In the absence of the long-range dipole-dipole interaction ($d=0$), the single-particle momentum distribution at high momenta is proportional to $1/q^4$, and the proportionality factor depends  on the derivative of the ground-sate energy with respect to the scattering length (Tan's adiabatic sweep theorem \cite{Tan2008b,Tan2008a}). The question arises how to extend Tan's adiabatic sweep theorem to the dipolar gases in three dimensions. This problem is a challenge for the theory of cold gases \cite{Hofmann21} because of the long-rang nature of the dipole-dipole interaction, which makes it impossible to apply Tan's theorem directly to these systems. In this paper, we suggest a solution of the problem by means of the variational theorem for the scattering length \cite{cherny00,cherny21,cherny22}, extended to the dipole-dipole interaction.

We found analytically the single-particle momentum distribution at high momenta, see Eq.~(\ref{Cherrelgen}). It is anisotropic and still proportional to the derivative of the ground-sate energy with respect to the scattering length. In addition, the short-range behaviour of the pair distribution function (\ref{grgen}) was obtained explicitly, which allows us to calculate the dynamic structure factor (\ref{ssfgen}) at large momenta. The equations (\ref{Cherrelgen}), (\ref{grgen}), and (\ref{ssfgen}) are the main results of the paper.

We emphasize that the tail of single-particle momentum distribution is calculated only for universal systems. If the three-body Efimov states for the dipole-dipole interaction \cite{wang11} give a significant contribution to the ground-state energy then an additional control parameter should appear by analogy with radially-symmetric short-range potentials \cite{werner08,werner09,*werner12}.

The paper is organized as follows. The next section regards the dipole-dipole interaction and its properties. In particular, the behaviour of the two-body scattering at small distances are considered, and the formula for the short-range scattering amplitude is derived. Besides, an analytical formula is found for the zero-energy scattering amplitude. In Sec.~\ref{sec:varth}, we obtain the variational theorem for the scattering length in the presence of the dipole-dipole interaction. In Sec.~\ref{results}, the relations are found for the single-particle momentum distribution, pair distribution function, and static structure factor. Finally, the obtained results are summarized and some prospects are discussed in Sec.~\ref{sec:concl}.

\section{The dipole-dipole interaction and its properties}

\subsection{The dipole-dipole interaction}
\label{sec:pseudo}

A realistic interaction in the dipolar Bose gas can be presented as the sum
\begin{align}\label{fullint}
V(\bm{r})=V_{0}(r)+V_{\mathrm{dd}}(\bm{r},{l_0})
\end{align}
of a short-range potential $V_{0}(r)$, typically decreasing at large distances as $1/r^6$ \cite{Baranov12}, and the long-range dipole tail
\begin{align}\label{ddintcut}
 V_{\mathrm{dd}}(\bm{r},{l_0})=\begin{cases}
 -2d^2 P_2(\hat{d}\cdot\hat{r})/{r^3} ,  &\text{for } r\geqslant {l_0},\\
 0, &\text{for } r < {l_0}.
 \end{cases}
\end{align}
Here ${l_0}$ is a cutoff parameter of order of atomic size\footnote{Note that in our paper \cite{cherny19b}, the notation $r_0$ was used instead of $l_0$.}, $P_2(x)=(3x^2-1)/2$ is the Legendre polynomial of the second order and $\hat{r}= \bm{r}/r$ is the unit vectors along the direction of the relative coordinate $\bm{r}$. The dipoles are supposed to be parallel and aligned along the direction $\hat{d}$  by a homogeneous external field. We denote the characteristic radius of the short-range interaction $V_{0}(r)$ as $r_0$. If the short-range potential contains a repulsive core at distances more than $l_0$ then one can put $l_0=0$ in Eq.~(\ref{ddintcut}).

The Fourier transform of the long-range part of the interaction (\ref{ddintcut}) in the limit $l_0\to0$ is given by
\begin{align}\label{vddqr0}
V_{\mathrm{dd}}(\bm{q})=\frac{8\pi d^2}{3} P_2(\hat{d}\cdot\hat{q}),
\end{align}
which is independent of the absolute value of $\bm{q}$.

At low energies, the form of the short-range potential $V_{0}(r)$ is not important, and its physical properties is controlled by the low-energy scattering amplitude, which is proportional to the scattering length $a$ (universality). This enables us to introduce the pseudopotential, which implies, to put it simply, that the short-range part of a real two-body interaction (\ref{fullint}) is replaced by $\frac{4\pi\hbar^2 a}{m}\delta(\bm{r})$. Universality for cold dipolar gases assumes that two potentials of different shape are indistinguishable at the low-energy scale as long as they have the same scattering length and the effective dipole range. It follows that the ground-state energy depends only on the particle mass, scattering length, effective dipole range, and parameters of an external trap.

\subsection{The wave function of two oriented dipoles at small distances}

The scattering amplitude $f$ for a short-range potential is defined \cite{llvol3_77} through the long-range asymptotics of the wavefunction
\begin{align}\label{sc_ampl}
\varphi(\bm{r})\simeq e^{i\bm{p}\cdot\bm{r}} + f\frac{e^{ipr}}{r},
\end{align}
which obeys the Schr\"odinger equation
\begin{align}\label{SchrEqfull}
(\nabla^{2}+p^2)\varphi =2m^*V(\bm{r})\varphi/\hbar^2.
\end{align}
Here $\bm{p}$ is the wavevector of incident particles of mass $m^{*}$, which are scattered by the potential $V(\bm{r})$. In the case of two-body scattering with the relative momentum $\hbar\bm{p}$, we should put $m^{*}=m/2$.

The potential (\ref{fullint}) is of long-range type due to the dipole part proportional to $P_2(\hat{d}\cdot\hat{r})/{r^3}$. Nevertheless, the plane wave at large distances is not distorted, and the scattering part of the wavefunction remains  proportional to $1/r$. Thus the asymptotics (\ref{sc_ampl}) remains intact. However, the long-range nature of the dipole-dipole interaction leads to a complicated behaviour of $f$. In particular, it differs \emph{qualitatively} in the ranges $r\gg 1/p$ and $r\ll 1/p$.

The scattering amplitude $f$ is obtained from the asymptotics (\ref{sc_ampl}) in the regime $r\gg 1/p$. It depends on \emph{all} the directions involved: $\hat{d}$, $\hat{r}$, and $\hat{p}$. This can be seen even in the Born approximation (see Appendix \ref{sec:Born}).

On the other hand, at the distances $r\ll 1/p$, one can go the limit $p=0$ in Eq.~(\ref{SchrEqfull}). Then the asymptotics of its solution in the range $r_0\ll r\ll 1/p$ takes the form \cite{cherny19b}
\begin{align}\label{Atot}
\varphi(\bm{r})\simeq 1
-\frac{a-r_\mathrm{dd}P_2(\hat{d}\cdot\hat{r})}{r}.
\end{align}
The relation (\ref{Atot}) can be checked directly with the identities $\nabla^2 1/r=0$ and $\nabla^2 P_2(\cos\vartheta)/{r}=-6 P_2(\cos\vartheta)/{r^3}$ for arbitrary $r\not=0$. Here $\cos\vartheta = \hat{d}\cdot\hat{r}$ when the $z$-axis is parallel to the direction of the dipole moment. Thus the
%long-distance
tail of the dipole-dipole interaction in the Schr\"odinger equation cancels out due to the presence of the anisotropic part of the wavefunction $r_\mathrm{dd}P_2(\cos\vartheta)/{r}$.

One can also call the quantity  $f=-a +r_\mathrm{dd}P_2(\hat{d}\cdot\hat{r})$ in Eq.~(\ref{Atot}) the ``short-range'' scattering amplitude. It is \emph{independent} of the directions of incident particles $\hat{p}$. We show in the Born approximation (see Appendix \ref{sec:Born}) that the small-momentum correction to $f$ is proportional to $pr$ with the prefactor depending on the direction $\hat{p}$. This agrees well with the above results.

\subsection{The scattering amplitude and pseudopotential for the dipole-diploe interaction}
\label{pseudopot}

To find the wavefunction in the range $r_0\ll r\ll 1/p$, it is sufficient to consider the two-body scattering at zero energy. This implies that $p=0$ and $m^{*}=m/2$ in the Schr\"odinger equation (\ref{SchrEqfull}). We are looking for an even solution of this equation.
%in the form $\varphi(\bm{r}) =\sum_{l=0}^{\infty}\varphi_{2l}(r)P_{2l}(\hat{d}\cdot\hat{r})$ with $P_{l}$ being the Legendre polynomials.
Separating the scattering part $\psi(\bm{r})$ of the wave function $\varphi(\bm{r})=1+\psi(\bm{r})$ and taking the Fourier transformation of the Schr\"odinger equation yield
\begin{align}
\psi(\bm{q})=-\frac{U(\bm{q})}{2T({\bm{q}})}.
\label{SchFour}
\end{align}
Here $T({\bm{q}})=\hbar^2q^2/(2m)$ and $\psi(\bm{q})$ are the free-particle dispersion and the Fourier transform of $\psi(\bm{r})$, respectively, and
\begin{align}\label{Uqgen}
 U(\bm{q}) =\int {d}r^3\, e^{-i\bm{q}\cdot\bm{r}}V(\bm{r})\varphi(\bm{r})
\end{align}
is the scattering amplitude "off the mass shell"\footnote{For a short-range potential, the scattering amplitude $f$ in Eq.~(\ref{sc_ampl}) is proportional to the matrix element $\int {d}r^3\, e^{-i\bm{q}\cdot\bm{r}}V(\bm{r})\varphi_{\bm{p}}(\bm{r})$, where $\varphi_{\bm{p}}(\bm{r})$ is the solution of the Schr\"oinger equation (\ref{SchrEqfull}) on the "mass shell" $p=q$. For simplicity, we also call $U(\bm{q})$ given by Eq.~(\ref{Uqgen}) the scattering amplitude.}. To obtain the scattering amplitude, we need to know the solution $\varphi(\bm{r})$.

For the wavefunction, the range $r_0\ll r\ll 1/p$ in real space corresponds to $p\ll q\ll 1/r_0$ in momentum space. Then, in Eq.~(\ref{SchFour}), we can approximate the scattering amplitude by its low-momentum asymptotics: $U(\bm{q})\simeq U_\mathrm{ps}(\bm{q})$ . Expanding the wavefunction into partial waves with the Legendre polynomials
\begin{align}\label{phiexp}
 \varphi(\bm{r}) =\sum_{l=0}^{\infty}\varphi_{2l}(r)P_{2l}(\hat{d}\cdot\hat{r})
\end{align}
and substituting the expansion into Eq.~(\ref{Uqgen}), we obtain the low-energy scattering amplitude \cite{Schutzhold06} for $p\ll q\ll 1/r_0$
\begin{align}\label{uqeffps}
U_\mathrm{ps}(\bm{q})=&\frac{4\pi\hbar^2 a}{m} +\frac{8\pi d^2}{3}P_2(\hat{d}\cdot\hat{q})\nonumber\\
=&\frac{4\pi\hbar^2 a}{m}\left[1-\epsilon_{\mathrm{dd}}+3 \epsilon_{\mathrm{dd}}(\hat{d}\cdot\hat{q})^2\right],
\end{align}
where the scattering length is given by \cite{cherny19b}
\begin{align}\label{aeq}
a= \frac{m}{\hbar^2}\int_{0}^{\infty}\!\! {d}r\, r^2 V_{0}(r)\varphi_{0}(r)
                         -\frac{6r_{\mathrm{dd}}}{5}\int_{l_{0}}^{\infty} {d}r \frac{\varphi_{2}(r)}{r}.
\end{align}
The low-energy scattering amplitude (\ref{uqeffps}) is independent of the absolute value of the wavevector but depends on its direction. For this reason,  its value at $q=0$ is undefined. The low-momentum amplitude (\ref{uqeffps}) matches well with the long-range behaviour (\ref{Atot}) of the wavefunction, which are related through Eq.~(\ref{SchFour}).

Equation (\ref{Atot}) tells us that two lowest components $\varphi_{0}(r)\simeq1-a/r$ and $\varphi_{2}(r)\simeq r_{\mathrm{dd}}/r$  explicitly determine the main asymptotic behaviour of the wave function when $r\to\infty$. However, the Schr\"odinger equation relates any $\varphi_{2l}$ component to  $\varphi_{2l-2}$ and $\varphi_{2l+2}$ and thus the partial waves do not separate, as opposed to the case of a radially symmetric potential (see the detailed discussion in Ref.~\cite{cherny19b}). This implies that all even momenta implicitly contribute to the scattering length.

The inverse Fourier transformation of the low-energy scattering amplitude gives us the effective pseudopotential for the dipolar gas \cite{Yi00}
\begin{align}
V_\mathrm{ps}(\bm{r})=\!\int {d}q^3\, e^{i\bm{q}\cdot\bm{r}}U_\mathrm{ps}(\bm{q})=\frac{4\pi\hbar^2 a}{m}\delta(\bm{r})\!-\!\frac{2d^2}{r^3}P_2(\hat{d}\cdot\hat{r}).\nonumber
\end{align}

\subsection{The analytical expression for the zero-energy scattering amplitude}
\label{scat_ampl_large}

We define the scattering amplitude $a$ through the asymptotics (\ref{Atot}) of the Schr\"odinger equation (\ref{SchrEqfull}) with $p=0$. As explained above, this asymptotics corresponds to the \emph{short-range} behaviour of the solution of the Schr\"odinger equation with $p\not=0$ in the range $r_0\ll r\ll 1/p$. On the other hand, the scattering length is usually defined in the literature \cite{Yi00,Ronen06} as the spherically symmetric component of the \emph{long-range} scattering amplitude [see Eq.~(\ref{sc_ampl})] in the zero-energy limit $p\to0$. Let us show that these definitions are equivalent and obtain analytically the zero-energy scattering amplitude as a function of the incident $\hat{p}$ and scattered $\hat{r}$ directions and the scattering length $a$.

The scattering amplitude can be expanded in the spherical harmonics \cite{Yi00}
\begin{align}\label{fharm}
f(\hat{p},\hat{r},p)=4\pi\sum_{lm,l'm'}t_{lm}^{l'm'}(p)Y_{lm}^{*}(\hat{p})Y_{l'm'}(\hat{r})
\end{align}
with $t_{lm}^{l'm'}(p)=\langle l'm' |\bm{T}|lm\rangle/p$ being the reduced $T$-matrix elements. As was found in Ref.~\cite{Yi00}, all even momenta make finite contributions to the scattering amplitude in the zero-energy limit $p\to0$. By definition, the scattering length is given by the  \emph{spherically symmetric} component of the reduced $T$-matrix in this limit: $a=-\lim_{p\to 0}t_{00}^{00}(p)$. Then we obtain from Eq.~(\ref{fharm})
\begin{align}
a=-\lim_{p\to 0}\langle f(\hat{p},\hat{r},p)\rangle_{\hat{p},\hat{r}},\label{adeft}
%-t_{00}^{00}(0) =-\frac{1}{4\pi}\int d\hat{p}\,d\hat{r} f(\hat{p},\hat{r},0) Y_{00}(\hat{p})Y_{00}^{*}(\hat{r})\nonumber\\
\end{align}
where the brackets $\langle\cdots\rangle_{\hat{p},\hat{r}} =\frac{1}{(4\pi)^2}\int d\hat{p}\,d\hat{r}\cdots $ stand for the average over all directions of unit vectors $\hat{p}$ and $\hat{r}$.

The scattering amplitude can be written down through the matrix element (see the discussion in Appendix \ref{sec:Born})
\begin{align}
f(\hat{p},\hat{r},p)=&-\frac{m}{4\pi\hbar^2}U_{p\hat{p}}(p\hat{r}),\label{amplmatr}\\
U_{\bm{p}}(\bm{q})=&\int d^3y\, e^{-i\bm{q}\cdot\bm{y}} V(\bm{y}) \varphi_{\bm{p}}(\bm{y})\label{matrel}
\end{align}
with $\varphi_{\bm{p}}(\bm{y})$ being the solution of the Schr\"odinger equation (\ref{SchrEqfull}). Separating the plane wave $\varphi_{\bm{p}}(\bm{y})=e^{i\bm{p}\cdot\bm{y}}+\psi_{\bm{p}}(\bm{y})$ and substituting $\varphi_{\bm{p}}(\bm{y})$ into Eq.~(\ref{matrel}) yield
\begin{align}\label{UpUqp}
 U_{\bm{p}}(\bm{q})=&\int d^3y\, e^{-i(\bm{q}-\bm{p})\cdot\bm{y}} V_{\mathrm{dd}}(\bm{y})+\int d^3y\, e^{-i\bm{q}\cdot\bm{y}} V_{0}(y)\varphi_{\bm{p}}(\bm{y})\nonumber\\
 &+\int d^3y\, e^{-i\bm{q}\cdot\bm{y}} V_{\mathrm{dd}}(\bm{y})\psi_{\bm{p}}(\bm{y}),
\end{align}
where $\bm{p}=p\hat{p}$ and $\bm{q}=p\hat{r}$.

The first term in Eq.~(\ref{UpUqp}) is calculated explicitly. In the limit $p\to0$, it is given by the Fourier transform (\ref{vddqr0}) of $\hat{r}-\hat{p}$. The second and third integrals converge absolutely for large $y$, because the short-range potential $V_{0}(y)$ falls off faster than $1/y^3$ and $|V_{\mathrm{dd}}(\bm{y})\psi_{\bm{p}}(\bm{y})|\sim 1/y^4$ when $y\to\infty$. Then we can go to the limit $p\to 0$ under the integrals, which yields $\int d^3y\,V_{0}(y)\varphi(\bm{y})$ and $\int d^3y V_{\mathrm{dd}}(\bm{y})\psi(\bm{y})$ for the second and third terms, respectively. Here $\varphi(\bm{y})=1+\psi(\bm{y})$ is the solution of the the Schr\"odinger equation (\ref{SchrEqfull}) at $p=0$, because for any fixed $\bm{y}$ and $p\to0$ we arrive at the short-range regime $yp\ll 1$. Using the expansion (\ref{phiexp}) and  the orthogonality relations for the Legendre polynomials $\int d\hat{y} P_{l}(\hat{d}\cdot\hat{y})P_{l'}(\hat{d}\cdot\hat{y}) =4\pi\delta_{ll'}/(2l+1)$,
it is easy to verify that the sum of the the second and third terms amounts to $4\pi\hbar^2a/m$, where the scattering length is given by
Eq.~(\ref{aeq}). Then we obtain from Eq.~(\ref{amplmatr}) in the zero-energy limit
\begin{align}
  \lim_{p\to0}f(\hat{p},\hat{r},p)=-a-2r_{\mathrm{dd}}P_{2}\left(\frac{\hat{d}\cdot(\hat{r}-\hat{p})}{\sqrt{2(1-\hat{p}\cdot\hat{r})}}\right) \label{f0p}
\end{align}
with the scattering amplitude (\ref{aeq}).

The average of the last term of Eq.~(\ref{f0p}) with respect to all directions of unit vectors $\hat{p}$ and $\hat{r}$ is equal to zero, and we are left with Eq.~(\ref{adeft}). This proves the equivalence of the different definitions of the scattering length.

The reduced $T$-matrix elements at $p\to0$ could be obtained analytically from Eqs.~(\ref{fharm}) and (\ref{f0p}) through the corresponding  Clebsch-Gordan coefficients. However, this hardly makes sense, since the full scattering  amplitude is already known explicitly in elementary functions. We simply note that the obtained expression for the zero-energy scattering amplitude (\ref{f0p}) reveals that all even momenta make finite contributions to it due to the presence of the Born term of the long-range dipole-dipole interaction. This conclusion is consistent with numerical computations of Ref.~\cite{Ronen06}.

\section{The variational theorem for the scattering length in the presence of the dipole-dipole interaction}
\label{sec:varth}

The variational theorem for the scattering length was suggested in our previous publications \cite{cherny00,cherny21,cherny22}. Although it was obtained for short-range radially symmetric potentials, the theorem can easily be generalized to the dipole-dipole interaction, for which it takes a similar form. We apply the variational theorem for the total potential (\ref{fullint}).

The main problem of the generalization lies in the long-range nature of the dipole-dipole interaction, which leads to the logarithmic divergence of the integral $\int d^3r{V}(\bm{r})\varphi(\bm{r})$, in which the wavefunction $\varphi(\bm{r})$ obeys Eq.~(\ref{SchrEqfull}) at $p=0$ and has the asymptotics (\ref{Atot}). As a consequence, the limit of the short-range scattering amplitude (\ref{uqeffps}) as $q\to0$ does not exist and $U(0)$ is not defined. However, the above integral can be understood as the limit $\int d^3r\cdots =\lim_{R\to\infty}\int_{0}^{R} dr\, r^2 \int d\hat{r}\cdots$, where $\int d\hat{r}$ denotes the integration over all directions of the radius vector. Then the zero-momentum scattering amplitude is well-defined and given by
\begin{align}\label{scatamp}
U(0)=\int d^3r{V}(\bm{r})\varphi(\bm{r})=\frac{4\pi\hbar^2a}{m}.
\end{align}
This relation can be proved with the help of the Gauss-Ostrogradsky (divergence) theorem.

Following Ref.~\cite{cherny22}, we consider the two-particle scattering problem with zero total energy and momentum for arbitrary single-particle dispersion $T(\bm{q})$. It is supposed to be quite close to the usual free-particle dispersion $T({q})=\hbar^2q^2/(2m)$ (see Ref.~\cite{cherny22} for details). The two-body Schr\"odinger equation takes the form
\begin{equation}
2T\left(-i\hbar\nabla\right)\varphi(\bm{r})+{V}(\bm{r})\varphi(\bm{r})=0.\nonumber
\end{equation}

Separating the scattering part of the wave function $\varphi(\bm{r})=1+\psi(\bm{r})$ and following the same arguments as in the papers \cite{cherny21,cherny22}, we obtain the zero-momentum scattering amplitude
\begin{equation}
{U}^{}(0)
=\int{d}^3r\left[\psi(\bm{r})2T\left(-i\hbar\nabla\right)\psi(\bm{r})
                            + {V}(\bm{r})\varphi^2(\bm{r})\right]\nonumber
\end{equation}
and arrive at the variational theorem
\begin{align}
\delta{U}^{}(0)=\int\frac{{d}^3 q}{(2\pi)^3}\,
2\delta T(\bm{q})\psi^{2}(\bm{q})+\int{d}^3 r\,\delta{V}(\bm{r})\varphi^{2}(\bm{r}).\nonumber
%\label{varth}
\end{align}
Here the Fourier transform of the scattering part of the wave function is given by Eq.~(\ref{SchFour}). The last integral converges if it is understood as discussed above.

We are looking for the behaviour of the two-body wavefunction in the ranges $r_0\ll r\ll 1/p$ and  $p\ll q\ll 1/r_0$ in real and momentum spaces, respectively. Then one can replace the scattering amplitude by its low-momentum limit (\ref{uqeffps}) and the wavefunction by its asymptotics (\ref{Atot}), which finally yields
\begin{align}
  \frac{\delta U(0)}{\delta T(\bm{q})} =&\,2\psi^{2}(\bm{q})= \frac{1}{(2\pi)^3}\frac{U^2_\mathrm{ps}(\bm{q})}{2T^{2}(q)}
  =\frac{m^2U^2_\mathrm{ps}(\bm{q})}{4\pi^3\hbar^4q^4}, \label{varTp}\\
  \frac{\delta U(0)}{\delta V(\bm{r})} =&\, \varphi^{2}(\bm{r})=\left(1-\frac{a-r_\mathrm{dd}P_2(\hat{d}\cdot\hat{r})}{r}\right)^2.\label{varV}
\end{align}
The second equality in Eq.~(\ref{varTp}) assumes that the variation is taken in the vicinity of the usual free-particle dispersion.

\section{Applications of the variational theorem}
\label{results}

In general, the ground-state energy depends on the interparticle interactions, single-particle dispersion, and parameters of a trapping potential. Universality implies that it depends on the scattering length, dipole range, and frequency of the harmonic trap: $E=E(a,r_{\mathrm{dd}},\omega)$.  On the other hand, the momentum single-particle distribution can be calculated by varying the energy with respect to the single-particle dispersion \cite{cherny21}: $\frac{\delta E}{\delta T({\bm{q}})} =\frac{N({\bm{q}})}{(2\pi)^3}$. For a universal system, the single-particle dispersion is involved through the scattering length, which yields $\frac{\delta E}{\delta T({\bm{q}})}=\frac{\partial E}{\partial a} \frac{\delta a}{\delta T({\bm{q}})}$. Then the variational theorem allows us to obtain the momentum distribution at large momenta. Similar arguments can be used to calculate the spatial correlations.

\subsection{The long-range tail of the momentum distribution}
\label{momdistr}

\begin{figure}[!tb]
\centerline{\includegraphics[width=.45\columnwidth]{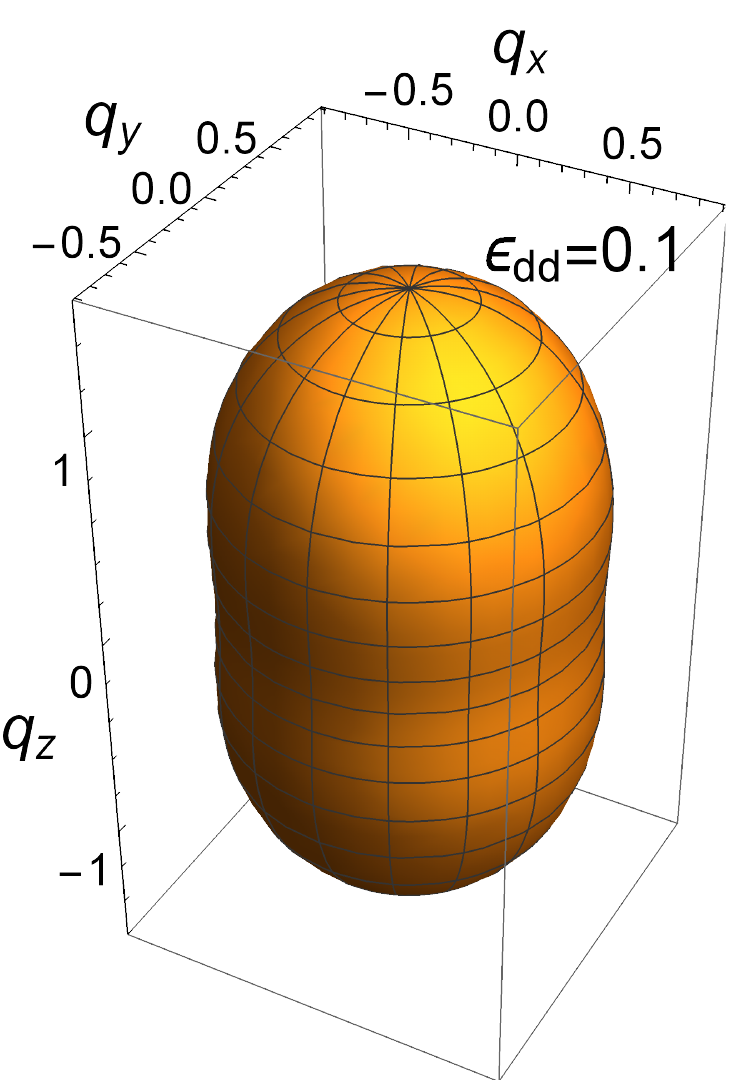}\ \includegraphics[width=.38\columnwidth]{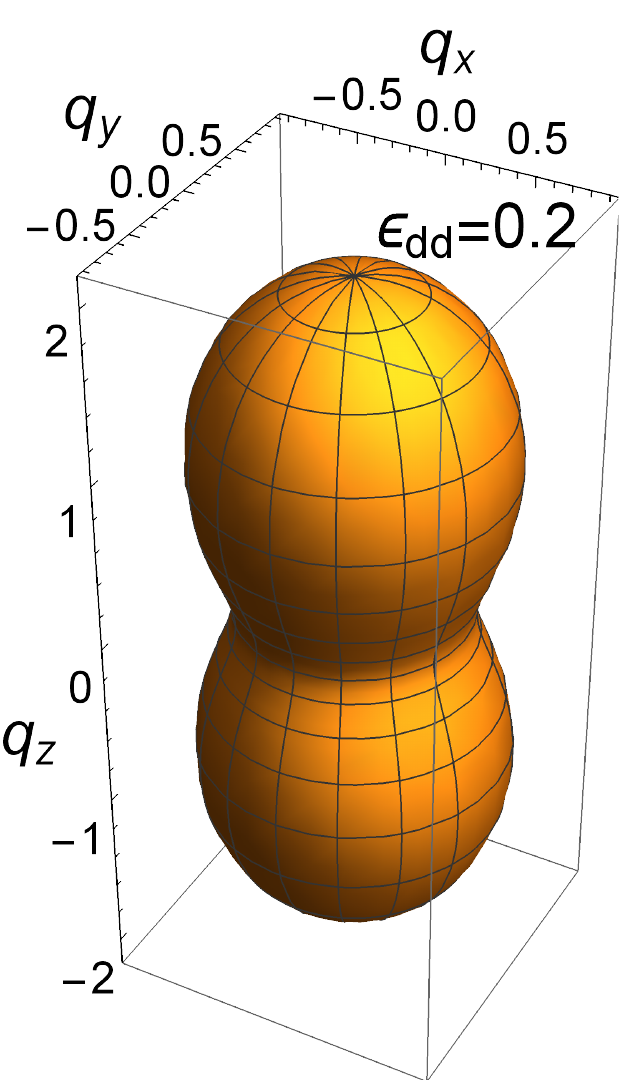}}
\centerline{\includegraphics[width=.42\columnwidth]{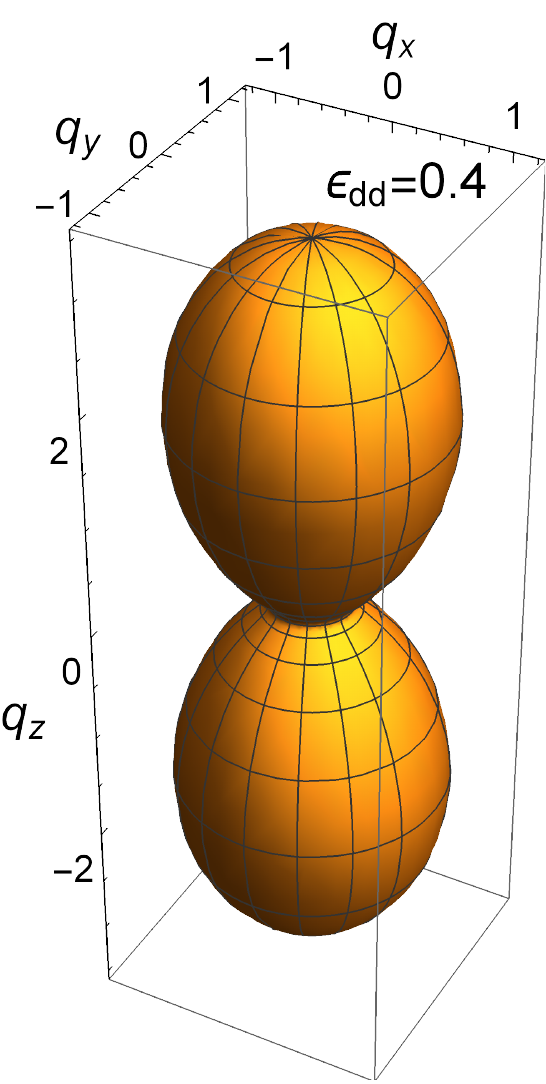}\ \ \ \includegraphics[width=.39\columnwidth]{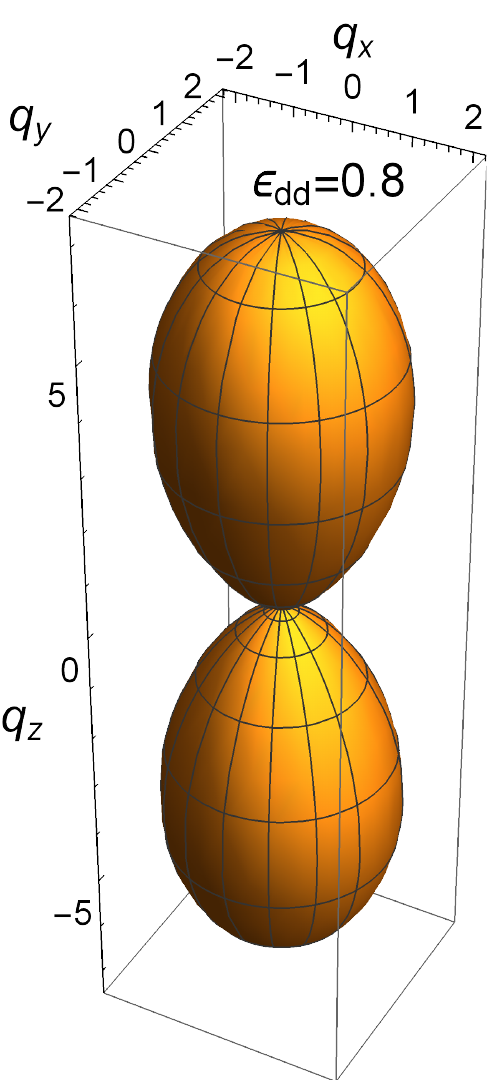}}
\caption{\label{fig:factor} Three-dimensional polar plot of the anisotropic factor $h(\cos\theta)$ [Eq.~(\ref{prefact})] in the single-particle momentum distribution (\ref{Cherrelgen}) for different values of the parameter $\epsilon_{\mathrm{dd}}$. The diagram shows spherical radius $h$ as a function of spherical coordinates $\theta$ and $\phi$: $q_x= h \sin\theta\cos\phi$, $q_y= h\sin\theta\sin\phi$, and $q_z= h\cos\theta$. The dipoles are aligned along the $z$-axis. The parameter $\epsilon_{\mathrm{dd}}$ is given by $\epsilon_{\mathrm{dd}}=r_{\mathrm{dd}}/a$ and characterises the relative strength of the dipole-dipole interaction. The condition $\epsilon_{\mathrm{dd}}<1$ is needed for stability of the dipolar Bose gas in three dimensions. The anisotropy of the momentum distribution strongly increases when $\epsilon_{\mathrm{dd}}$ gets closer to one.
 }
 \end{figure}

Using the above arguments, we can find the long-range tail of the single-particle momentum distribution in a trapped dipolar Bose-gas with the relation \cite{cherny21}
\begin{align}
N({\bm{q}})= (2\pi)^3\frac{\delta a}{\delta T({\bm{q}})}\left(\frac{\partial E}{\partial a}\right)_{r_{\mathrm{dd}},\omega,N},\nonumber
\end{align}
where $N$ is the total number of particles. For technical convenience, the mass is kept constant during the variation, because in this case Eq.~(\ref{scatamp}) yields $\delta U(0)=\frac{4\pi\hbar^2}{m}\delta a$, and besides $r_{\mathrm{dd}}$ and $\omega$ remains constant. With Eqs.~(\ref{uqeffps}) and (\ref{varTp}), we obtain
\begin{align}
N(\bm{q}) =&\,\frac{8\pi m a^2}{\hbar^2 q^4 }\left(\frac{\partial E}{\partial a}\right)_{r_{\mathrm{dd}},\omega,N} h(\hat{d}\cdot\hat{q}),
\label{Cherrelgen}\\
h(\hat{d}\cdot\hat{q})=&\left[1-\epsilon_{\mathrm{dd}}+3 \epsilon_{\mathrm{dd}}(\hat{d}\cdot\hat{q})^2\right]^{2}.
\label{prefact}
\end{align}
The distribution is  normalized to the total number of particles: $\frac{1}{(2\pi)^{3}}\int {d}^3 q\,N(\bm{q}) = N$. The scalar product of the unit vectors $\hat{d}\cdot\hat{q}$ amounts to $\cos\theta$ with $\theta$ being the angle between the directions of the dipoles and momentum. The prefactor obeys the inequality $(1-\epsilon_{\mathrm{dd}})^2 \leqslant h \leqslant (1+2\epsilon_{\mathrm{dd}})^2$. The stronger the dipole-dipole interaction, the bigger variation of the anisotropic factor $h$, and, therefore, the higher anisotropy of the momentum distribution, see Fig.~\ref{fig:factor}.

Near the unitary regime $a\to\infty$, the ground-state energy is linear with respect to $1/a$: $E=C_{0}-C_{1}/a+\cdots$ \cite{Pitaevskii16:book,werner12}. It follows that in this regime $a^2{\partial E}/{\partial a} =C_{1}$, and the left-hand-side of Eq.~(\ref{Cherrelgen}) remains finite.

For quantum gases, the typical value of de Broglie wavelength $2\pi/p$ should be of order of the mean distance between particles or higher. Then the relation (\ref{Cherrelgen}) is valid in the range $1/\xi\ll q\ll 1/r_0$, where $\xi$ determines the characteristic scale of the many-body effects, which is also of order of the mean distance.

In the particular case of the homogeneous gas, the distribution $N(\bm{q})$ is proportional to the mean occupation numbers $n_{\bm{q}}$: $N(\bm{q})=n_{\bm{q}}V$, where $V$ is the volume of the system. In Eq.~(\ref{Cherrelgen}), the partial derivative of the ground state with respect to the scattering length should be taken at constant volume.

The distribution over the absolute value of momentum is given by the average of Eq.~(\ref{Cherrelgen}) over the directions of $\hat{q}$, that is, over the solid angle with the volume element $d \hat{q} =\sin\vartheta d\varphi d\vartheta$:
\begin{align}
  N(q)&=\frac{1}{4\pi}\int d\hat{q}\,N(\bm{q})\nonumber \\
  &= \frac{8\pi m a^2}{\hbar^2q^4}\left(\frac{\partial E}{\partial a}\right)_{r_{\mathrm{dd}},\omega,N}
\left(1+\frac{4\epsilon_{\mathrm{dd}}^2}{5}\right).
\label{Cherrelgenav}
\end{align}

In the absence of the dipole-dipole interaction, $\epsilon_{\mathrm{dd}}=0$ Eqs.~(\ref{Cherrelgen}) and (\ref{Cherrelgenav}) are reduced to Tan's adiabatic sweep theorem
\begin{align}
  q^4 N({q})=&\,{\cal I}=\frac{8\pi m a^2}{\hbar^2}\left(\frac{\partial E}{\partial a}\right)_{\omega,N},\nonumber
\end{align}
where ${\cal I}$ is Tan's constant for trapped gases \cite{Tan2008b}. In this case, the long-range momentum distribution becomes radially symmetric.

The equation (\ref{Cherrelgenav}) cannot be applied for calculation of the kinetic energy per particle, which is proportional to the integral of $q^2 N(q)$ and thus diverges at large momenta. The reason is that the momentum distributions (\ref{Cherrelgen}) and (\ref{Cherrelgenav}) are valid in the range $1/\xi\ll q\ll 1/r_0$. In order to eliminate the divergence in the kinetic energy, we need to obtain the momentum distribution for $q\gtrsim 1/r_0$. Note that the same problem arises in the standard Tan's adiabatic sweep theorem \cite{Tan2008b,Tan2008a} for the same reason. In the papers \cite{cherny21,cherny22} by the author, the solution of this problem was suggested for a short-range potential. The same could be done for the dipole-dipole interaction. However, the range $q\gtrsim 1/r_0$ in the momentum space is not yet accessible experimentally, and in this paper we restrict ourselves to the range $1/\xi\ll q\ll 1/r_0$.

\subsection{The shot-range behaviour of the pair distribution function}
\label{pdf}

In this section, we find the short-range behaviour of the pair distribution function with the help of the variational theorem \cite{cherny21}. Here we restrict ourselves to the homogeneous case. The pair correlation function $g(\bm{r})$ determines the density-density correlations, and it is proportional to the conditional probability to find a particle at point $\bm{r}$ provided another particle is located at the origin $\bm{r}=0$.

The pair distribution function is obtained by means of variation of the ground-state with respect to the interaction potential %by analogy with the previous section
(see details in Ref.~\cite{cherny21})
\begin{align}
 g(\bm{r})= \frac{2}{n^2}\frac{1}{V}\frac{\delta E}{\delta V(\bm{r})}=\frac{2}{n^2}\frac{1}{V}\frac{\delta a}{\delta V({\bm{r}})}\left(\frac{\partial E}{\partial a}\right)_{r_{\mathrm{dd}},V,N},\nonumber
\end{align}
where $n=N/V$ is the density of particles. The variation $\delta{V}(\bm{r})$ is supposed to be localized within $r\ll \xi$, that is, it can be related to the short-range part of the interaction potential (\ref{fullint}): $\delta{V}(\bm{r})=\delta{V}_{0}(\bm{r})$. We get in this manner
\begin{align}
g(\bm{r})=\, &w\left[1- \frac{a-r_{\mathrm{dd}}P_2(\hat{d}\cdot\hat{r})}{r}\right]^2, \label{grgen}\\
w=\, &\frac{1}{V}\left(\frac{\partial E}{\partial a}\right)_{r_{\mathrm{dd}},V,N}\frac{m}{2\pi n^2\hbar^2}
%=\frac{{\cal C}}{16\pi^2a^2n^2}.
\label{factgr}
\end{align}
when $r_0\ll r\ll \xi$.

Note that we cannot use Eq.~(\ref{grgen}) for calculating the full interaction energy $\frac{N^2}{2V}\int d^3r\,g(\bm{r})V(\bm{r})$, because the contribution of the long-range dipole-dipole forces is significant at the distances of order $\xi$ and higher.

In the absence of the dipole-dipole interaction at $r_{\mathrm{dd}}$=0, we reproduce the results obtained by Cherny and Shanenko twenty years ago \cite{cherny00,cherny01}.

\subsection{The static structure factor}
\label{SSF}

For the homogeneous system, the static structure factor can be obtained (see Ref.~\cite{Pitaevskii16:book}, Sec.~18.5) by taking the Fourier transformation of the pair distribution function (\ref{grgen})
\begin{align}\label{sqgr}
S(\bm{q})=1+n\int{d}^3r[g(\bm{r})-1]e^{-i \bm{q}\cdot\bm{r}}.
\end{align}
In this way we get for $1/\xi\ll q\ll 1/r_0$
\begin{align}
S&(\bm{q})= 1-\frac{8\pi a n w}{q^2}\left(1-\epsilon_{\mathrm{dd}}+3\epsilon_{\mathrm{dd}}x^2\right)+\frac{\pi^2 a^2 n w}{q}\nonumber\\
&\times\left[2\!-\!\epsilon_{\mathrm{dd}}\!+\!\frac{11\epsilon_{\mathrm{dd}}^2}{16}\!+\!3\epsilon_{\mathrm{dd}}
\left(1\!-\!\frac{5\epsilon_{\mathrm{dd}}}{8}\right)x^2\!+\!\frac{27}{16}\epsilon_{\mathrm{dd}}^2x^4
\right],
\label{ssfgen}
\end{align}
where $q$ is the absolute value of $\bm{q}$, $w$ is given by Eq.~(\ref{factgr}), and we put by definition $x=\hat{d}\cdot\hat{q}=\cos\theta$.

Note that the $\delta$-function contribution, which  appears formally in the right-hand side of Eq.~(\ref{sqgr}), should be ignored. It arises at $w\not=1$ because  Eq.~(\ref{grgen}) does not describe correctly the long-range behaviour of the pair distribution function, which should tend to one when $r\gg \xi$.

\subsection{The weakly interacting spinless Bose gas at zero temperature}

We apply the obtained relations to the homogeneous dilute spinless Bose gas at zero temperature. Its ground-state energy is given by \cite{cherny19b}
\begin{align}
   E= \frac{2\pi\hbar^2a}{m}\frac{N^2}{V}\left[1-\epsilon_{\mathrm{dd}}+\frac{128}{15\sqrt{\pi}}\sqrt{n a^3}{\cal Q}_{5}(\epsilon_{\mathrm{dd}})\right], \label{expen}
\end{align}
where we use the notation
\begin{align}
{\cal Q}_{l}(y)=(1-y)^{l/2}{}_{2}F_{1}\left(\frac{1}{2},-\frac{l}{2};\frac{3}{2};-\frac{3y}{1-y}\right)\nonumber
%\label{Qn}
\end{align}
with ${}_{2}F_{1}$ being the hypergeometric function. Actually, Eq.~(\ref{expen}) is the ground-state expansion with respect to the small parameter $na^3$, and the stability condition requires that $a>0$ and $\epsilon_{\mathrm{dd}}=r_{\mathrm{dd}}/a<1$. At $\epsilon_{\mathrm{dd}}=0$, we arrive at the famous Lee-Huang-Yang result \cite{Lee57}, since ${\cal Q}_{5}(0)=1$.

The equation (\ref{expen}) yields
\begin{align}\label{dEda}
 \left(\frac{\partial E}{\partial a}\right)_{r_{\mathrm{dd}},V,N}=&\frac{2\pi\hbar^2}{m}\frac{N^2}{V}
\left[1+\frac{128}{15\sqrt{\pi}}\sqrt{n a^3}{\cal P}_{5}(\epsilon_{\mathrm{dd}})\right],\\
 {\cal P}_{l}(y)=&\frac{(l+1){\cal Q}_{l}(y)-(1+2y)^{l/2}}{2(1-y)}.\nonumber
\end{align}
Substituting Eq.~(\ref{dEda}) into the general relation (\ref{Cherrelgen}), we obtain the average occupation numbers when $1/\xi\ll q\ll 1/r_0$
\begin{align}\label{nqasymp}
 n_{\bm{q}}=\frac{16\pi^2n^2a^2}{q^4}f(\hat{d}\cdot\hat{q})\left[1+\frac{128}{15\sqrt{\pi}}\sqrt{n a^3}{\cal P}_{5}(\epsilon_{\mathrm{dd}})\right].
\end{align}

On the other hand, the Bogoliubov theory \cite{bogoliubov47} gives us for $\bm{q}\not=0$ \cite{cherny19b}
\begin{align}
n_{\bm{q}}=&\frac{1}{2}\left(\frac{T({q})+n U_\mathrm{ps}(\bm{q})}{\sqrt{T({q})^2+2n U_\mathrm{ps}(\bm{q})T({q})}}-1\right), \label{ocnum}
\end{align}
where $U_\mathrm{ps}(\bm{q})$ is the pseudopotential (\ref{uqeffps}) and $T({q})$ is the usual free-particle dispersion. The long-range tail of the occupation numbers (\ref{ocnum}) is given by
\begin{align}\label{nqasympB}
n_{\bm{q}}=\frac{1}{4}\frac{n^2 U^2_\mathrm{ps}(\bm{q})}{T^2({q})}=\frac{16\pi^2n^2a^2}{q^4}f(\hat{d}\cdot\hat{q}).
\end{align}
One can see that Eqs.~(\ref{nqasymp}) and (\ref{nqasympB}) coincide in the main order, but Eq.~(\ref{nqasymp}) is more precise and contains the correction of order of $\sqrt{na^3}$.

In the framework of the Bogoliubov theory, the dynamic structure factor takes the form
\begin{align}
S(\bm{q})=\frac{T({q})}{\sqrt{T({q})^2+2n U_\mathrm{ps}(\bm{q})T({q})}},\nonumber
\end{align}
which yields for $1/\xi\ll q\ll 1/r_0$
\begin{align}\label{SqasympB}
S(\bm{q})=1-\frac{n U_\mathrm{ps}(\bm{q})}{T({q})}=1-\frac{8\pi a n}{q^2}\left(1-\epsilon_{\mathrm{dd}}+3\epsilon_{\mathrm{dd}}x^2\right).
\end{align}
The accuracy of Eq.~(\ref{ssfgen}), obtained with the variational theorem, is higher. First, it contains the additional term proportional to $1/q$, and second, the coefficient $w$ in the first term includes the correction of order of $\sqrt{na^3}$:
\begin{align}\label{wB}
  w=1+\frac{128}{15\sqrt{\pi}}\sqrt{n a^3}{\cal P}_{5}(\epsilon_{\mathrm{dd}}),
\end{align}
as it follows from Eqs.~(\ref{factgr}) and (\ref{dEda}).

\section{Conclusions}
\label{sec:concl}

We extended the variational theorem for the scattering length \cite{cherny00,cherny21,cherny22} to the potential (\ref{fullint}), which is the sum of a short-range and dipole-dipole interactions. The generalization is based on the short-range asymptotics (\ref{Atot}) of the two-body wavefunction \cite{cherny19b}.

As a byproduct, we obtain analytical expression (\ref{f0p}) for the zero-energy scattering amplitude in the presence of the dipole-dipole interaction.

The variations of the ground-state energy with respect to the one-particle dispersion and the short-range part of the interaction enable us to obtain the single-particle momentum distribution (\ref{Cherrelgen}) at large momenta and the pair distribution function (\ref{grgen}) at small distances. For universal systems, these variations are realized through the scattering length, which allows us to apply the variational theorem (\ref{varTp}) and (\ref{varV}).

The momentum distribution contains the axially symmetric prefactor with the axis being parallel to the direction of the dipoles, see Fig.~\ref{fig:factor}. Integrating out the angles, we arrived at the distribution over the absolute value of momentum (\ref{Cherrelgenav}). We get Tan's adiabatic sweep theorem as a particular case of Eqs.~(\ref{Cherrelgen}) and  (\ref{Cherrelgenav}) when the dipole-dipole interaction is switched off.

The static structure factor (\ref{ssfgen}) was found with the Fourier transformation of the pair distribution function.

We verified the obtained general relations for the dipolar Bose gas in the particular case of the weakly interacting Bose gas \cite{cherny19b}, although the relations are applicable for arbitrary large values of the scattering length.

The relations (\ref{Cherrelgen}), (\ref{grgen}), and (\ref{ssfgen}) can easily be generalized to finite temperatures. It suffices to replace the ground-state energy in the equations with the free energy (see the details in Ref.~\cite{cherny21}). This is all the more important when the scattering length is large, because finite temperatures are needed \cite{Chevy16} for stability of the Bose gas in the large-gas-parameter regime $na^3 \gg 1$. Such regime with a large scattering length together with high electric dipole moment could be realized in Rydberg atoms \cite{Balewski16} or polar bosonic molecules \cite{Moses17}, although in these systems the regime of Bose condensation is not reached yet experimentally.

\appendix

\section{The Born approximation for the dipole-dipole interaction}
\label{sec:Born}

We consider the scattering of two spinless bosons in the center-of-mass system and therefore are looking for an even solution of the Schr\"odinger equation (\ref{SchrEqfull}) with the potential (\ref{ddintcut}) when $l_0\to0$.

The wavefunction can be written in the form $\varphi_{\bm{p}}(\bm{r}) = \cos(\bm{p}\cdot\bm{r}) +\psi_{\bm{p}}(\bm{r})$. In the Born approximation,  the term $\psi_{\bm{p}}(\bm{r})$ in the right-hand side of Eq.~(\ref{SchrEqfull}) is neglected. Taking the Fourier transformation of the Schr\"odinger equation yields
\[
(-q^2+p^2)\psi_{\bm{p}}(\bm{q})=\frac{m}{\hbar^2}V_{\bm{p}}(\bm{q})
\]
where $\psi_{\bm{p}}(\bm{q})$ is the is the Fourier transform of $\psi_{\bm{p}}(\bm{r})$, and
\begin{align}\label{Symdd}
V_{\bm{p}}(\bm{q})=\frac{V_{\mathrm{dd}}(\bm{p}+\bm{q})+V_{\mathrm{dd}}(\bm{p}-\bm{q})}{2}
\end{align}
is the symmetrized Fourier transform of the dipole-dipole interaction (\ref{vddqr0}).

After the inverse Fourier transformation we are left with
\begin{align}\label{ShrBorn}
 \psi_{\bm{p}}(\bm{r})= -\frac{m}{(2\pi)^3\hbar^2}\int d^3q\, e^{i\bm{q}\cdot\bm{r}}\frac{V_{\bm{p}}(\bm{q})}{q^2-p^2-i0},
\end{align}
where the passage round the pole  $q^2=p^2$ removes the ambiguity of the result (see Ref.~\cite{llvol3_77}, Sec.~130).

The long-range asymptotics for $r\gg 1/p$ can be obtained from Eq.~(\ref{ShrBorn}) in the standard way \cite{llvol3_77}:
\begin{align}\label{asympBorn}
\psi_{\bm{p}}(\bm{r})\simeq -\frac{m V_{\hat{p}}(\hat{r})}{4\pi\hbar^2}\frac{e^{ipr}}{r}.
\end{align}
Here we use the scaling property of the Fourier transform of the dipole-dipole interaction (\ref{Symdd}): $V_{\bm{p}}(p\hat{r})=V_{\hat{p}}(\hat{r})$. The scattering amplitude depends on all the orientations: the dipoles $\hat{d}$, wavevector $\hat{p}$ of the relative momentum, and radius vector $\hat{r}$.

To calculate the scattering part of the wavefunction (\ref{ShrBorn}) when $r\ll 1/p$, we represent it in the form
\begin{align}
\psi_{\bm{p}}(\bm{r})=& -\frac{m}{(2\pi)^3\hbar^2}\int d^3q\, e^{i\bm{q}\cdot\bm{r}}\frac{V_{\mathrm{dd}}(\hat{q})}{q^2}-\frac{m}{(2\pi)^3\hbar^2}\nonumber\\
  &\times\int d^3q\, e^{i\bm{q}\cdot\bm{r}} \left[\frac{V_{\hat{p}}(\bm{q}/p)}{q^2-p^2-i0}-\frac{V_{\mathrm{dd}}(\hat{q})}{q^2}\right].\nonumber
\end{align}
The first integral is easily calculated as $r_\mathrm{dd}P_2(\hat{d}\cdot\hat{r})/r$. By means of the scaling substitution $\bm{q}=p\bm{y}$ in the second integral, we obtain in the limit $pr\to0$
\begin{align}
\psi_{\bm{p}}(\bm{r})\simeq&\frac{r_\mathrm{dd}P_2(\hat{d}\cdot\hat{r})}{r} - p\,C,\nonumber\\
C=&\frac{m}{(2\pi)^3\hbar^2}\int d^3y\left[\frac{V_{\hat{p}}(y\hat{y})}{y^2-1-i0}-\frac{V_{\mathrm{dd}}(\hat{y})}{y^2}\right]\nonumber.
\end{align}
The last integral converges, because $V_{\hat{p}}(y\hat{y})-V_{\mathrm{dd}}(\hat{y})=O(1/y^2)$ at large $y$. Here the symbol $O(z)$ denotes the terms of order of $z$ or smaller. One can see that the prefactor $C$ depends on the direction of the relative momentum $\hat{p}$, as expected.

The Born approximation for the dipole-dipole interaction was used in the literature many years ago \cite{Cross66}.

Note that beyond the Born approximation, Eq.~(\ref{ShrBorn}) is still valid but the Fourier transform of the dipole-dipole interaction $V_{\bm{p}}(\bm{q})$ should be replaced by the matrix element $U_{\bm{p}}(\bm{q})=\int d^3r e^{-i\bm{q}\cdot\bm{r}} V_{\mathrm{dd}}(\bm{r}) \varphi_{\bm{p}}(\bm{r})$. The matrix element is the sum of the Born term and the integral containing the scattering piece. The latter converges absolutely for large $r$, since the integrand is proportional to $1/r^4$, and then no problem arises with the scattering piece. With the same considerations, we arrive at the asymptotics (\ref{asympBorn}) proportional to $U_{\bm{p}}(p\hat{r})e^{ipr}/r$. This makes the scheme consistent: starting from the assumption that the asymptotics of the wavefunction $\varphi_{\bm{p}}(\bm{r})$ is the sum of the plane wave and the spherical wave proportional to $e^{ipr}/r$, we derive this asymptotics from the Schr\"odinger equation in momentum representation.

\bibliography{tanvar}

\begin{thebibliography}{30}%
\makeatletter
\providecommand \@ifxundefined [1]{%
 \@ifx{#1\undefined}
}%
\providecommand \@ifnum [1]{%
 \ifnum #1\expandafter \@firstoftwo
 \else \expandafter \@secondoftwo
 \fi
}%
\providecommand \@ifx [1]{%
 \ifx #1\expandafter \@firstoftwo
 \else \expandafter \@secondoftwo
 \fi
}%
\providecommand \natexlab [1]{#1}%
\providecommand \enquote  [1]{``#1''}%
\providecommand \bibnamefont  [1]{#1}%
\providecommand \bibfnamefont [1]{#1}%
\providecommand \citenamefont [1]{#1}%
\providecommand \href@noop [0]{\@secondoftwo}%
\providecommand \href [0]{\begingroup \@sanitize@url \@href}%
\providecommand \@href[1]{\@@startlink{#1}\@@href}%
\providecommand \@@href[1]{\endgroup#1\@@endlink}%
\providecommand \@sanitize@url [0]{\catcode `\\12\catcode `\$12\catcode
  `\&12\catcode `\#12\catcode `\^12\catcode `\_12\catcode `\%12\relax}%
\providecommand \@@startlink[1]{}%
\providecommand \@@endlink[0]{}%
\providecommand \url  [0]{\begingroup\@sanitize@url \@url }%
\providecommand \@url [1]{\endgroup\@href {#1}{\urlprefix }}%
\providecommand \urlprefix  [0]{URL }%
\providecommand \Eprint [0]{\href }%
\providecommand \doibase [0]{https://doi.org/}%
\providecommand \selectlanguage [0]{\@gobble}%
\providecommand \bibinfo  [0]{\@secondoftwo}%
\providecommand \bibfield  [0]{\@secondoftwo}%
\providecommand \translation [1]{[#1]}%
\providecommand \BibitemOpen [0]{}%
\providecommand \bibitemStop [0]{}%
\providecommand \bibitemNoStop [0]{.\EOS\space}%
\providecommand \EOS [0]{\spacefactor3000\relax}%
\providecommand \BibitemShut  [1]{\csname bibitem#1\endcsname}%
\let\auto@bib@innerbib\@empty
%</preamble>
\bibitem [{\citenamefont {Baranov}(2008)}]{Baranov08}%
  \BibitemOpen
  \bibfield  {author} {\bibinfo {author} {\bibfnamefont {M.~A.}\ \bibnamefont
  {Baranov}},\ }\bibfield  {title} {\bibinfo {title} {Theoretical progress in
  many-body physics with ultracold dipolar gases},\ }\href
  {https://doi.org/10.1016/j.physrep.2008.04.007} {\bibfield  {journal}
  {\bibinfo  {journal} {Phys. Rep.}\ }\textbf {\bibinfo {volume} {464}},\
  \bibinfo {pages} {71} (\bibinfo {year} {2008})}\BibitemShut {NoStop}%
\bibitem [{\citenamefont {Lahaye}\ \emph {et~al.}(2009)\citenamefont {Lahaye},
  \citenamefont {Menotti}, \citenamefont {Santos}, \citenamefont {Lewenstein},\
  and\ \citenamefont {Pfau}}]{Lahaye09}%
  \BibitemOpen
  \bibfield  {author} {\bibinfo {author} {\bibfnamefont {T.}~\bibnamefont
  {Lahaye}}, \bibinfo {author} {\bibfnamefont {C.}~\bibnamefont {Menotti}},
  \bibinfo {author} {\bibfnamefont {L.}~\bibnamefont {Santos}}, \bibinfo
  {author} {\bibfnamefont {M.}~\bibnamefont {Lewenstein}},\ and\ \bibinfo
  {author} {\bibfnamefont {T.}~\bibnamefont {Pfau}},\ }\bibfield  {title}
  {\bibinfo {title} {The physics of dipolar bosonic quantum gases},\ }\href
  {https://doi.org/10.1088/0034-4885/72/12/126401} {\bibfield  {journal}
  {\bibinfo  {journal} {Rep. Prog. Phys.}\ }\textbf {\bibinfo {volume} {72}},\
  \bibinfo {pages} {126401} (\bibinfo {year} {2009})}\BibitemShut {NoStop}%
\bibitem [{\citenamefont {Baranov}\ \emph {et~al.}(2012)\citenamefont
  {Baranov}, \citenamefont {Dalmonte}, \citenamefont {Pupillo},\ and\
  \citenamefont {Zoller}}]{Baranov12}%
  \BibitemOpen
  \bibfield  {author} {\bibinfo {author} {\bibfnamefont {M.~A.}\ \bibnamefont
  {Baranov}}, \bibinfo {author} {\bibfnamefont {M.}~\bibnamefont {Dalmonte}},
  \bibinfo {author} {\bibfnamefont {G.}~\bibnamefont {Pupillo}},\ and\ \bibinfo
  {author} {\bibfnamefont {P.}~\bibnamefont {Zoller}},\ }\bibfield  {title}
  {\bibinfo {title} {Condensed matter theory of dipolar quantum gases},\ }\href
  {https://doi.org/10.1021/cr2003568} {\bibfield  {journal} {\bibinfo
  {journal} {Chem. Rev.}\ }\textbf {\bibinfo {volume} {112}},\ \bibinfo {pages}
  {5012} (\bibinfo {year} {2012})}\BibitemShut {NoStop}%
\bibitem [{\citenamefont {Pitaevskii}\ and\ \citenamefont
  {Stringari}(2016)}]{Pitaevskii16:book}%
  \BibitemOpen
  \bibfield  {author} {\bibinfo {author} {\bibfnamefont {L.}~\bibnamefont
  {Pitaevskii}}\ and\ \bibinfo {author} {\bibfnamefont {S.}~\bibnamefont
  {Stringari}},\ }\href
  {https://doi.org/10.1093/acprof:oso/9780198758884.001.0001} {\emph {\bibinfo
  {title} {{B}ose-{E}instein Condensation and Superfluidity}}}\ (\bibinfo
  {publisher} {Oxford University},\ \bibinfo {address} {Oxford},\ \bibinfo
  {year} {2016})\BibitemShut {NoStop}%
\bibitem [{\citenamefont {Ferrier-Barbut}(2019)}]{Ferrier-Barbut19}%
  \BibitemOpen
  \bibfield  {author} {\bibinfo {author} {\bibfnamefont {I.}~\bibnamefont
  {Ferrier-Barbut}},\ }\bibfield  {title} {\bibinfo {title} {Ultradilute
  quantum droplets},\ }\href {https://doi.org/10.1063/PT.3.4184} {\bibfield
  {journal} {\bibinfo  {journal} {Phys. Today}\ }\textbf {\bibinfo {volume}
  {72}},\ \bibinfo {pages} {46} (\bibinfo {year} {2019})}\BibitemShut {NoStop}%
\bibitem [{\citenamefont {Luo}\ \emph {et~al.}(2021)\citenamefont {Luo},
  \citenamefont {Pang}, \citenamefont {Liu}, \citenamefont {Li},\ and\
  \citenamefont {Malomed}}]{Luo20}%
  \BibitemOpen
  \bibfield  {author} {\bibinfo {author} {\bibfnamefont {Z.-H.}\ \bibnamefont
  {Luo}}, \bibinfo {author} {\bibfnamefont {W.}~\bibnamefont {Pang}}, \bibinfo
  {author} {\bibfnamefont {B.}~\bibnamefont {Liu}}, \bibinfo {author}
  {\bibfnamefont {Y.-Y.}\ \bibnamefont {Li}},\ and\ \bibinfo {author}
  {\bibfnamefont {B.~A.}\ \bibnamefont {Malomed}},\ }\bibfield  {title}
  {\bibinfo {title} {A new form of liquid matter: Quantum droplets},\ }\href
  {https://doi.org/10.1007/s11467-020-1020-2} {\bibfield  {journal} {\bibinfo
  {journal} {Front. Phys.}\ }\textbf {\bibinfo {volume} {16}},\ \bibinfo
  {pages} {32201} (\bibinfo {year} {2021})}\BibitemShut {NoStop}%
\bibitem [{\citenamefont {B\"ottcher}\ \emph {et~al.}(2020)\citenamefont
  {B\"ottcher}, \citenamefont {Schmidt}, \citenamefont {Hertkorn},
  \citenamefont {Ng}, \citenamefont {Graham}, \citenamefont {Guo},
  \citenamefont {Langen},\ and\ \citenamefont {Pfau}}]{Bottcher20}%
  \BibitemOpen
  \bibfield  {author} {\bibinfo {author} {\bibfnamefont {F.}~\bibnamefont
  {B\"ottcher}}, \bibinfo {author} {\bibfnamefont {J.-N.}\ \bibnamefont
  {Schmidt}}, \bibinfo {author} {\bibfnamefont {J.}~\bibnamefont {Hertkorn}},
  \bibinfo {author} {\bibfnamefont {K.~S.~H.}\ \bibnamefont {Ng}}, \bibinfo
  {author} {\bibfnamefont {S.~D.}\ \bibnamefont {Graham}}, \bibinfo {author}
  {\bibfnamefont {M.}~\bibnamefont {Guo}}, \bibinfo {author} {\bibfnamefont
  {T.}~\bibnamefont {Langen}},\ and\ \bibinfo {author} {\bibfnamefont
  {T.}~\bibnamefont {Pfau}},\ }\bibfield  {title} {\bibinfo {title} {New states
  of matter with fine-tuned interactions: quantum droplets and dipolar
  supersolids},\ }\href {https://doi.org/10.1088/1361-6633/abc9ab} {\bibfield
  {journal} {\bibinfo  {journal} {Rep. Prog. Phys.}\ }\textbf {\bibinfo
  {volume} {84}},\ \bibinfo {pages} {012403} (\bibinfo {year}
  {2020})}\BibitemShut {NoStop}%
\bibitem [{\citenamefont {Cherny}(2019)}]{cherny19b}%
  \BibitemOpen
  \bibfield  {author} {\bibinfo {author} {\bibfnamefont {A.~{\relax Yu}.}\
  \bibnamefont {Cherny}},\ }\bibfield  {title} {\bibinfo {title} {Low-density
  expansions for the homogeneous dipolar {B}ose gas at zero temperature},\
  }\href {https://doi.org/10.1103/PhysRevA.100.063631} {\bibfield  {journal}
  {\bibinfo  {journal} {Phys. Rev. A}\ }\textbf {\bibinfo {volume} {100}},\
  \bibinfo {pages} {063631} (\bibinfo {year} {2019})}\BibitemShut {NoStop}%
\bibitem [{\citenamefont {Tan}(2008{\natexlab{a}})}]{Tan2008b}%
  \BibitemOpen
  \bibfield  {author} {\bibinfo {author} {\bibfnamefont {S.}~\bibnamefont
  {Tan}},\ }\bibfield  {title} {\bibinfo {title} {Large momentum part of a
  strongly correlated {F}ermi gas},\ }\href
  {https://doi.org/https://doi.org/10.1016/j.aop.2008.03.005} {\bibfield
  {journal} {\bibinfo  {journal} {Ann. Phys.}\ }\textbf {\bibinfo {volume}
  {323}},\ \bibinfo {pages} {2971} (\bibinfo {year}
  {2008}{\natexlab{a}})}\BibitemShut {NoStop}%
\bibitem [{\citenamefont {Tan}(2008{\natexlab{b}})}]{Tan2008a}%
  \BibitemOpen
  \bibfield  {author} {\bibinfo {author} {\bibfnamefont {S.}~\bibnamefont
  {Tan}},\ }\bibfield  {title} {\bibinfo {title} {Energetics of a strongly
  correlated {F}ermi gas},\ }\href
  {https://doi.org/https://doi.org/10.1016/j.aop.2008.03.004} {\bibfield
  {journal} {\bibinfo  {journal} {Ann. Phys.}\ }\textbf {\bibinfo {volume}
  {323}},\ \bibinfo {pages} {2952} (\bibinfo {year}
  {2008}{\natexlab{b}})}\BibitemShut {NoStop}%
\bibitem [{\citenamefont {Hofmann}\ and\ \citenamefont
  {Zwerger}(2021)}]{Hofmann21}%
  \BibitemOpen
  \bibfield  {author} {\bibinfo {author} {\bibfnamefont {J.}~\bibnamefont
  {Hofmann}}\ and\ \bibinfo {author} {\bibfnamefont {W.}~\bibnamefont
  {Zwerger}},\ }\bibfield  {title} {\bibinfo {title} {Universal relations for
  dipolar quantum gases},\ }\href
  {https://doi.org/10.1103/PhysRevResearch.3.013088} {\bibfield  {journal}
  {\bibinfo  {journal} {Phys. Rev. Research}\ }\textbf {\bibinfo {volume}
  {3}},\ \bibinfo {pages} {013088} (\bibinfo {year} {2021})}\BibitemShut
  {NoStop}%
\bibitem [{\citenamefont {Cherny}\ and\ \citenamefont
  {Shanenko}(2000)}]{cherny00}%
  \BibitemOpen
  \bibfield  {author} {\bibinfo {author} {\bibfnamefont {A.~{\relax Yu}.}\
  \bibnamefont {Cherny}}\ and\ \bibinfo {author} {\bibfnamefont {A.~A.}\
  \bibnamefont {Shanenko}},\ }\bibfield  {title} {\bibinfo {title} {Short-range
  particle correlations in a dilute {B}ose gas},\ }\href
  {https://doi.org/10.1103/PhysRevE.62.1646} {\bibfield  {journal} {\bibinfo
  {journal} {Phys. Rev. E}\ }\textbf {\bibinfo {volume} {62}},\ \bibinfo
  {pages} {1646} (\bibinfo {year} {2000})}\BibitemShut {NoStop}%
\bibitem [{\citenamefont {Cherny}(2021)}]{cherny21}%
  \BibitemOpen
  \bibfield  {author} {\bibinfo {author} {\bibfnamefont {A.~{\relax Yu}.}\
  \bibnamefont {Cherny}},\ }\bibfield  {title} {\bibinfo {title} {Tan's
  adiabatic sweep theorem from the variational theorem for the scattering
  length},\ }\href {https://doi.org/10.1103/PhysRevA.104.043304} {\bibfield
  {journal} {\bibinfo  {journal} {Phys. Rev. A}\ }\textbf {\bibinfo {volume}
  {104}},\ \bibinfo {pages} {043304} (\bibinfo {year} {2021})}\BibitemShut
  {NoStop}%
\bibitem [{\citenamefont {Cherny}(2022)}]{cherny22}%
  \BibitemOpen
  \bibfield  {author} {\bibinfo {author} {\bibfnamefont {A.~{\relax Yu}.}\
  \bibnamefont {Cherny}},\ }\bibfield  {title} {\bibinfo {title} {The
  variational theorem for the scattering length in low dimensions and its
  applications to universal systems},\ }\href
  {https://doi.org/10.1088/1751-8121/ac57d0} {\bibfield  {journal} {\bibinfo
  {journal} {J. Phys. A: Math. Theor.}\ }\textbf {\bibinfo {volume} {55}},\
  \bibinfo {pages} {155004} (\bibinfo {year} {2022})}\BibitemShut {NoStop}%
\bibitem [{\citenamefont {Wang}\ \emph {et~al.}(2011)\citenamefont {Wang},
  \citenamefont {D'Incao},\ and\ \citenamefont {Greene}}]{wang11}%
  \BibitemOpen
  \bibfield  {author} {\bibinfo {author} {\bibfnamefont {Y.}~\bibnamefont
  {Wang}}, \bibinfo {author} {\bibfnamefont {J.~P.}\ \bibnamefont {D'Incao}},\
  and\ \bibinfo {author} {\bibfnamefont {C.~H.}\ \bibnamefont {Greene}},\
  }\bibfield  {title} {\bibinfo {title} {Efimov effect for three interacting
  bosonic dipoles},\ }\href {https://doi.org/10.1103/PhysRevLett.106.233201}
  {\bibfield  {journal} {\bibinfo  {journal} {Phys. Rev. Lett.}\ }\textbf
  {\bibinfo {volume} {106}},\ \bibinfo {pages} {233201} (\bibinfo {year}
  {2011})}\BibitemShut {NoStop}%
\bibitem [{\citenamefont {Werner}(2008)}]{werner08}%
  \BibitemOpen
  \bibfield  {author} {\bibinfo {author} {\bibfnamefont {F.}~\bibnamefont
  {Werner}},\ }\bibfield  {title} {\bibinfo {title} {Virial theorems for
  trapped cold atoms},\ }\href {https://doi.org/10.1103/PhysRevA.78.025601}
  {\bibfield  {journal} {\bibinfo  {journal} {Phys. Rev. A}\ }\textbf {\bibinfo
  {volume} {78}},\ \bibinfo {pages} {025601} (\bibinfo {year}
  {2008})}\BibitemShut {NoStop}%
\bibitem [{\citenamefont {Werner}\ \emph {et~al.}(2009)\citenamefont {Werner},
  \citenamefont {Tarruell},\ and\ \citenamefont {Castin}}]{werner09}%
  \BibitemOpen
  \bibfield  {author} {\bibinfo {author} {\bibfnamefont {F.}~\bibnamefont
  {Werner}}, \bibinfo {author} {\bibfnamefont {L.}~\bibnamefont {Tarruell}},\
  and\ \bibinfo {author} {\bibfnamefont {Y.}~\bibnamefont {Castin}},\
  }\bibfield  {title} {\bibinfo {title} {Number of closed-channel molecules in
  the {BEC-BCS} crossover},\ }\href
  {https://doi.org/10.1140/epjb/e2009-00040-8} {\bibfield  {journal} {\bibinfo
  {journal} {Eur. Phys. J. B}\ }\textbf {\bibinfo {volume} {68}},\ \bibinfo
  {pages} {401} (\bibinfo {year} {2009})}\BibitemShut {NoStop}%
\bibitem [{\citenamefont {Werner}\ and\ \citenamefont
  {Castin}(2012)}]{werner12}%
  \BibitemOpen
  \bibfield  {author} {\bibinfo {author} {\bibfnamefont {F.}~\bibnamefont
  {Werner}}\ and\ \bibinfo {author} {\bibfnamefont {Y.}~\bibnamefont
  {Castin}},\ }\bibfield  {title} {\bibinfo {title} {General relations for
  quantum gases in two and three dimensions. {II}. {B}osons and mixtures},\
  }\href {https://doi.org/10.1103/PhysRevA.86.053633} {\bibfield  {journal}
  {\bibinfo  {journal} {Phys. Rev. A}\ }\textbf {\bibinfo {volume} {86}},\
  \bibinfo {pages} {053633} (\bibinfo {year} {2012})}\BibitemShut {NoStop}%
\bibitem [{\citenamefont {Landau}\ and\ \citenamefont
  {Lifshitz}(1977)}]{llvol3_77}%
  \BibitemOpen
  \bibfield  {author} {\bibinfo {author} {\bibfnamefont {L.~D.}\ \bibnamefont
  {Landau}}\ and\ \bibinfo {author} {\bibfnamefont {E.~M.}\ \bibnamefont
  {Lifshitz}},\ }\href@noop {} {\emph {\bibinfo {title} {Quantum {M}echanics
  ({N}on-relativistic {T}heory)}}},\ \bibinfo {edition} {3rd}\ ed.,\ {C}ourse
  of {T}heoretical {P}hysics, {V}ol. 3\ (\bibinfo  {publisher} {Pergamon},\
  \bibinfo {address} {N.Y.},\ \bibinfo {year} {1977})\BibitemShut {NoStop}%
\bibitem [{\citenamefont {Sch\"utzhold}\ \emph {et~al.}(2006)\citenamefont
  {Sch\"utzhold}, \citenamefont {Uhlmann}, \citenamefont {Xu},\ and\
  \citenamefont {Fischer}}]{Schutzhold06}%
  \BibitemOpen
  \bibfield  {author} {\bibinfo {author} {\bibfnamefont {R.}~\bibnamefont
  {Sch\"utzhold}}, \bibinfo {author} {\bibfnamefont {M.}~\bibnamefont
  {Uhlmann}}, \bibinfo {author} {\bibfnamefont {Y.}~\bibnamefont {Xu}},\ and\
  \bibinfo {author} {\bibfnamefont {U.~R.}\ \bibnamefont {Fischer}},\
  }\bibfield  {title} {\bibinfo {title} {Mean-field expansion in
  {B}ose-{E}instein condensates with finite-range interactions},\ }\href
  {https://doi.org/10.1142/S0217979206035631} {\bibfield  {journal} {\bibinfo
  {journal} {Int. J. Mod. Phys. B}\ }\textbf {\bibinfo {volume} {20}},\
  \bibinfo {pages} {3555} (\bibinfo {year} {2006})}\BibitemShut {NoStop}%
\bibitem [{\citenamefont {Yi}\ and\ \citenamefont {You}(2000)}]{Yi00}%
  \BibitemOpen
  \bibfield  {author} {\bibinfo {author} {\bibfnamefont {S.}~\bibnamefont
  {Yi}}\ and\ \bibinfo {author} {\bibfnamefont {L.}~\bibnamefont {You}},\
  }\bibfield  {title} {\bibinfo {title} {Trapped atomic condensates with
  anisotropic interactions},\ }\href
  {https://doi.org/10.1103/PhysRevA.61.041604} {\bibfield  {journal} {\bibinfo
  {journal} {Phys. Rev. A}\ }\textbf {\bibinfo {volume} {61}},\ \bibinfo
  {pages} {041604(R)} (\bibinfo {year} {2000})}\BibitemShut {NoStop}%
\bibitem [{\citenamefont {Ronen}\ \emph {et~al.}(2006)\citenamefont {Ronen},
  \citenamefont {Bortolotti}, \citenamefont {Blume},\ and\ \citenamefont
  {Bohn}}]{Ronen06}%
  \BibitemOpen
  \bibfield  {author} {\bibinfo {author} {\bibfnamefont {S.}~\bibnamefont
  {Ronen}}, \bibinfo {author} {\bibfnamefont {D.~C.~E.}\ \bibnamefont
  {Bortolotti}}, \bibinfo {author} {\bibfnamefont {D.}~\bibnamefont {Blume}},\
  and\ \bibinfo {author} {\bibfnamefont {J.~L.}\ \bibnamefont {Bohn}},\
  }\bibfield  {title} {\bibinfo {title} {Dipolar {B}ose-{E}instein condensates
  with dipole-dependent scattering length},\ }\href
  {https://doi.org/10.1103/PhysRevA.74.033611} {\bibfield  {journal} {\bibinfo
  {journal} {Phys. Rev. A}\ }\textbf {\bibinfo {volume} {74}},\ \bibinfo
  {pages} {033611} (\bibinfo {year} {2006})}\BibitemShut {NoStop}%
\bibitem [{\citenamefont {Cherny}\ and\ \citenamefont
  {Shanenko}(2001)}]{cherny01}%
  \BibitemOpen
  \bibfield  {author} {\bibinfo {author} {\bibfnamefont {A.~{\relax Yu}.}\
  \bibnamefont {Cherny}}\ and\ \bibinfo {author} {\bibfnamefont {A.~A.}\
  \bibnamefont {Shanenko}},\ }\bibfield  {title} {\bibinfo {title} {Dilute
  {B}ose gas: short-range particle correlations and ultraviolet divergence},\
  }\href {https://doi.org/10.1007/s100510170301} {\bibfield  {journal}
  {\bibinfo  {journal} {Eur. Phys. J. B}\ }\textbf {\bibinfo {volume} {19}},\
  \bibinfo {pages} {555} (\bibinfo {year} {2001})}\BibitemShut {NoStop}%
\bibitem [{\citenamefont {Lee}\ \emph {et~al.}(1957)\citenamefont {Lee},
  \citenamefont {Huang},\ and\ \citenamefont {Yang}}]{Lee57}%
  \BibitemOpen
  \bibfield  {author} {\bibinfo {author} {\bibfnamefont {T.~D.}\ \bibnamefont
  {Lee}}, \bibinfo {author} {\bibfnamefont {K.}~\bibnamefont {Huang}},\ and\
  \bibinfo {author} {\bibfnamefont {C.~N.}\ \bibnamefont {Yang}},\ }\bibfield
  {title} {\bibinfo {title} {Eigenvalues and eigenfunctions of a {B}ose system
  of hard spheres and its low-temperature properties},\ }\href
  {https://doi.org/10.1103/PhysRev.106.1135} {\bibfield  {journal} {\bibinfo
  {journal} {Phys. Rev.}\ }\textbf {\bibinfo {volume} {106}},\ \bibinfo {pages}
  {1135} (\bibinfo {year} {1957})}\BibitemShut {NoStop}%
\bibitem [{\citenamefont {Bogoliubov}(1947)}]{bogoliubov47}%
  \BibitemOpen
  \bibfield  {author} {\bibinfo {author} {\bibfnamefont {N.~N.}\ \bibnamefont
  {Bogoliubov}},\ }\bibfield  {title} {\bibinfo {title} {On the theory of
  superfluidity},\ }\href@noop {} {\bibfield  {journal} {\bibinfo  {journal}
  {J. Phys. USSR}\ }\textbf {\bibinfo {volume} {11}},\ \bibinfo {pages} {23}
  (\bibinfo {year} {1947})},\ \bibinfo {note} {reprinted in
  Ref.~\cite{pines61:book}}\BibitemShut {NoStop}%
\bibitem [{\citenamefont {Chevy}\ and\ \citenamefont
  {Salomon}(2016)}]{Chevy16}%
  \BibitemOpen
  \bibfield  {author} {\bibinfo {author} {\bibfnamefont {F.}~\bibnamefont
  {Chevy}}\ and\ \bibinfo {author} {\bibfnamefont {C.}~\bibnamefont
  {Salomon}},\ }\bibfield  {title} {\bibinfo {title} {Strongly correlated
  {B}ose gases},\ }\href {https://doi.org/10.1088/0953-4075/49/19/192001}
  {\bibfield  {journal} {\bibinfo  {journal} {J. Phys. B: At. Mol. Opt. Phys.}\
  }\textbf {\bibinfo {volume} {49}},\ \bibinfo {pages} {192001} (\bibinfo
  {year} {2016})}\BibitemShut {NoStop}%
\bibitem [{\citenamefont {Balewski}\ and\ \citenamefont
  {Pfau}(2016)}]{Balewski16}%
  \BibitemOpen
  \bibfield  {author} {\bibinfo {author} {\bibfnamefont {J.}~\bibnamefont
  {Balewski}}\ and\ \bibinfo {author} {\bibfnamefont {T.}~\bibnamefont
  {Pfau}},\ }\bibinfo {title} {Spectroscopy of {R}ydberg atoms in dense
  ultracold gases},\ in\ \href {https://doi.org/10.3254/978-1-61499-694-1-443}
  {\emph {\bibinfo {booktitle} {Quantum Matter at Ultralow Temperatures}}},\
  \bibinfo {series} {Proceedings of the International School of Physics "Enrico
  Fermi"}, Vol.\ \bibinfo {volume} {191},\ \bibinfo {editor} {edited by\
  \bibinfo {editor} {\bibfnamefont {W.}~\bibnamefont {Ketterle}}, \bibinfo
  {editor} {\bibfnamefont {M.}~\bibnamefont {Inguscio}}, \bibinfo {editor}
  {\bibfnamefont {G.}~\bibnamefont {Roati}},\ and\ \bibinfo {editor}
  {\bibfnamefont {S.}~\bibnamefont {Stringari}}}\ (\bibinfo  {publisher} {IOS
  Press},\ \bibinfo {address} {Amsterdam},\ \bibinfo {year} {2016})\ pp.\
  \bibinfo {pages} {443--462}\BibitemShut {NoStop}%
\bibitem [{\citenamefont {Moses}\ \emph {et~al.}(2017)\citenamefont {Moses},
  \citenamefont {Covey}, \citenamefont {Miecnikowski}, \citenamefont {Jin},\
  and\ \citenamefont {Ye}}]{Moses17}%
  \BibitemOpen
  \bibfield  {author} {\bibinfo {author} {\bibfnamefont {S.}~\bibnamefont
  {Moses}}, \bibinfo {author} {\bibfnamefont {J.}~\bibnamefont {Covey}},
  \bibinfo {author} {\bibfnamefont {M.}~\bibnamefont {Miecnikowski}}, \bibinfo
  {author} {\bibfnamefont {D.}~\bibnamefont {Jin}},\ and\ \bibinfo {author}
  {\bibfnamefont {J.}~\bibnamefont {Ye}},\ }\bibfield  {title} {\bibinfo
  {title} {New frontiers for quantum gases of polar molecules},\ }\href
  {https://doi.org/10.1038/nphys3985} {\bibfield  {journal} {\bibinfo
  {journal} {Nat. Phys.}\ }\textbf {\bibinfo {volume} {13}},\ \bibinfo {pages}
  {13} (\bibinfo {year} {2017})}\BibitemShut {NoStop}%
\bibitem [{\citenamefont {Cross}\ and\ \citenamefont {Gordon}(1966)}]{Cross66}%
  \BibitemOpen
  \bibfield  {author} {\bibinfo {author} {\bibfnamefont {R.~J.}\ \bibnamefont
  {Cross}}\ and\ \bibinfo {author} {\bibfnamefont {R.~G.}\ \bibnamefont
  {Gordon}},\ }\bibfield  {title} {\bibinfo {title} {Long--range scattering
  from anisotropic potentials: Dipole--dipole scattering},\ }\href
  {https://doi.org/10.1063/1.1727375} {\bibfield  {journal} {\bibinfo
  {journal} {J. Chem. Phys.}\ }\textbf {\bibinfo {volume} {45}},\ \bibinfo
  {pages} {3571} (\bibinfo {year} {1966})}\BibitemShut {NoStop}%
\bibitem [{\citenamefont {Pines}(1990)}]{pines61:book}%
  \BibitemOpen
  \bibinfo {editor} {\bibfnamefont {D.}~\bibnamefont {Pines}},\ ed.,\
  \href@noop {} {\emph {\bibinfo {title} {The {M}any-{B}ody {P}roblem}}}\
  (\bibinfo  {publisher} {Benjamin},\ \bibinfo {address} {N. Y.},\ \bibinfo
  {year} {1990})\BibitemShut {NoStop}%
\end{thebibliography}%

\end{document}